\RequirePackage{fix-cm} 
\documentclass[smallextended]{svjour3} 
\smartqed 
\usepackage{graphicx} 
\journalname{Hyperfine Interactions} 

\begin{document} 

\title{Pion-assisted $N\Delta$ and $\Delta\Delta$ dibaryons, 
and beyond{\footnote{Invited talk at EXA 2014, Vienna, Sept. 2014, published 
in Hyperfine Interactions (2015) DOI 10.1007/s10751-015-1133-0}}} 
\titlerunning{Pion-assisted dibaryons} 

\author{Avraham Gal, Jerusalem}  
\authorrunning{A.~Gal} 
\institute{A. Gal \at Racah Institute of Physics, The Hebrew University, 
Jerusalem~91904, Israel \\ \email{avragal@savion.huji.ac.il}} 

\date{Received: date / Accepted: date} 

\maketitle 

\begin{abstract} 
Experimental evidence for $I(J^P)$=$0(3^+)$ $\Delta\Delta$ dibaryon 
${\cal D}_{03}(2370)$ has been presented recently by the WASA-at-COSY 
Collaboration. Here I review new hadronic-basis Faddeev calculations of $L=0$ 
nonstrange pion-assisted $N\Delta$ and $\Delta\Delta$ dibaryon candidates. 
These calculations are so far the only ones to reproduce the relatively small 
${\cal D}_{03}(2370)$ width of 70--80~MeV. Predictions are also given for the 
location and width of ${\cal D}_{30}$, the $I(J^P)$=$3(0^+)$ exotic partner 
of ${\cal D}_{03}(2370)$. Extensions to strangeness $\cal S$=$-$1 dibaryons 
are briefly discussed.  


\end{abstract}

\section{Introduction} 
\label{sec:intro} 

\begin{figure*}[htb]  
\includegraphics[width=0.48\textwidth,height=4.5cm]{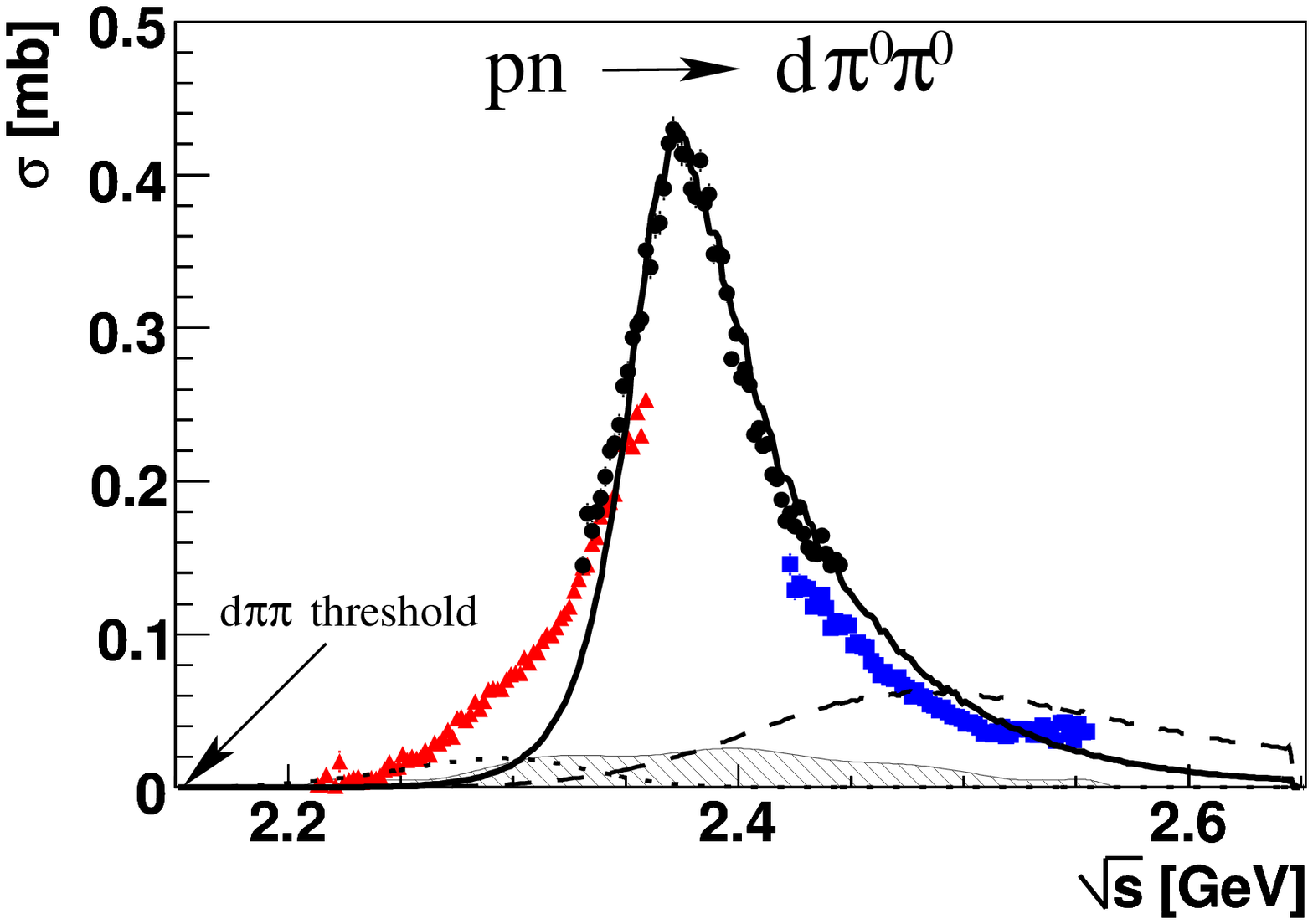} 
\includegraphics[width=0.48\textwidth,height=4.5cm]{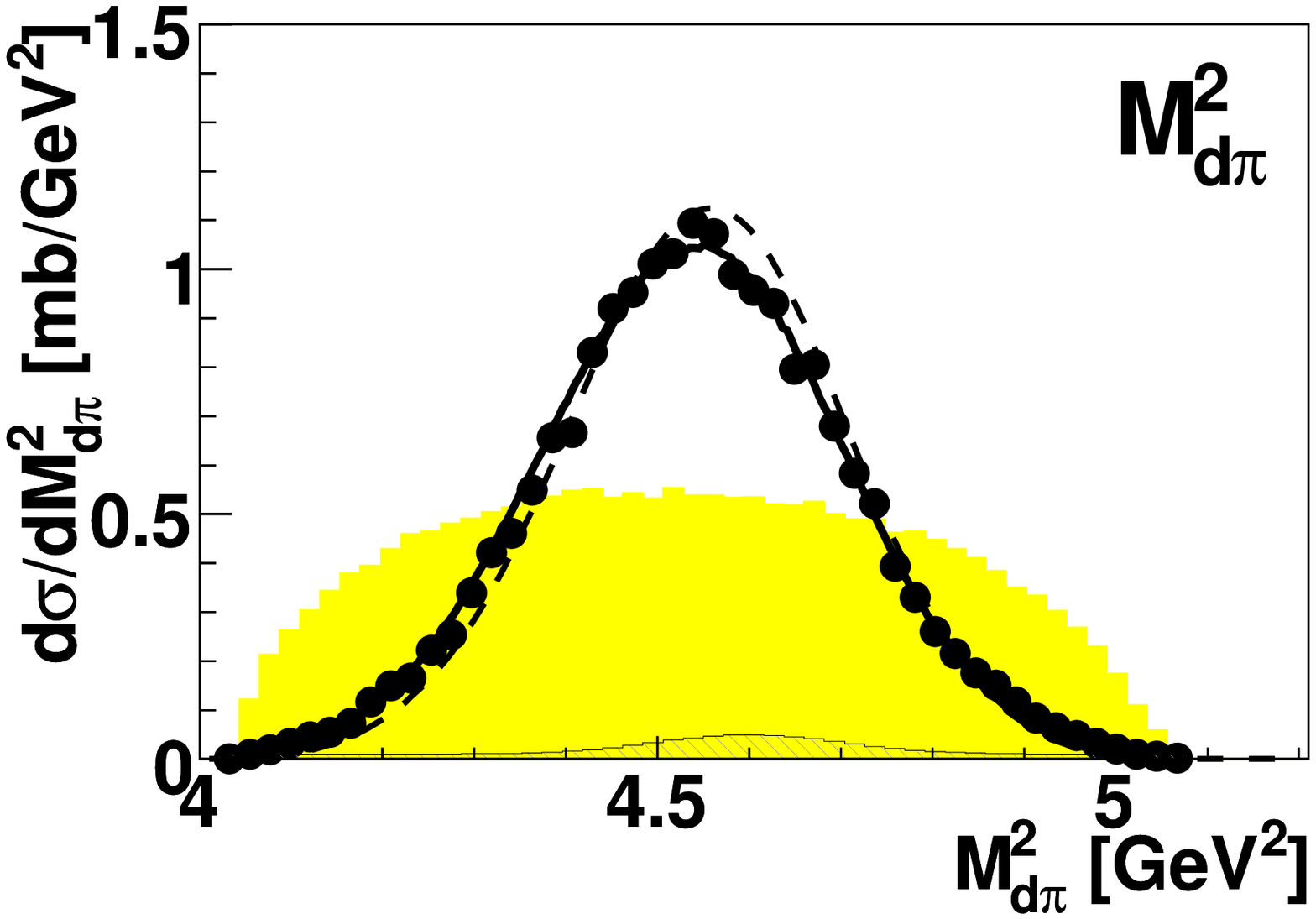} 
\caption{${\cal D}_{03}(2370)$ $\Delta\Delta$ dibaryon resonance signal on 
the left panel, and its $M^2_{d\pi^0}$ Dalitz-plot projection on the right 
panel, from $pn\to d\pi^0\pi^0$ measurements by WASA-at-COSY \cite{wasa11}. 
This resonance was also observed consistently in $pn\to d\pi^+\pi^-$ 
measurements \cite{wasa13}. Figures courtesy of Heinz Clement.} 
\label{fig:data} 
\end{figure*}

The WASA-at-COSY Collaboration has presented recently striking evidence 
for a $I(J^P)=0(3^+)$ nonstrange dibaryon resonance some 80--90 MeV below 
$2M_{\Delta}\simeq 2.46$~GeV, with a relatively small width of $\Gamma\approx 
70-80$~MeV, by observing a distinct resonance in $pn\to d\pi\pi$ reactions 
\cite{wasa11,wasa13} as shown in Fig.~\ref{fig:data}, left panel. 
Isospin $I=0$ is uniquely fixed in this particular $\pi^0\pi^0$ production 
reaction and the spin-parity $3^+$ assignment follows from the measured 
deuteron and pions angular distributions, assuming $s$-wave $\Delta\Delta$ 
decaying pair. The shape of the $M^2_{d\pi}$ distribution on the right panel 
supports $\Delta\Delta$ assignment and its peak at $\sqrt{s}\approx 2.13$~GeV, 
almost at the ${\cal D}_{12}(2150)$ $N\Delta$ dibaryon location (see below), 
might suggest a possible role for ${\cal D}_{12}$ in forming the 
$\Delta\Delta$ dibaryon ${\cal D}_{03}$. 

\begin{figure*}[htb]  
\includegraphics[width=0.48\textwidth,height=4.5cm]{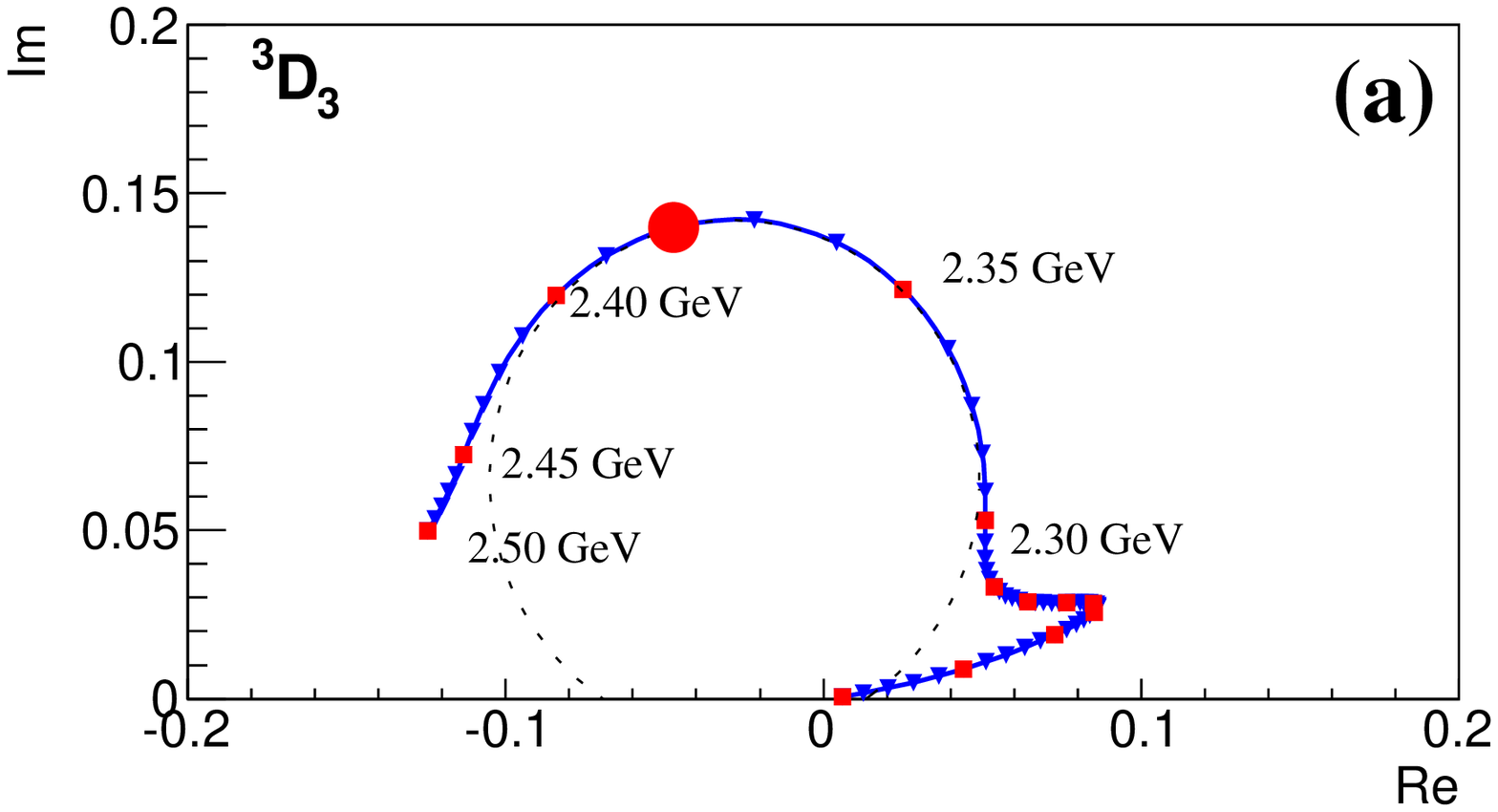} 
\includegraphics[width=0.48\textwidth,height=4.5cm]{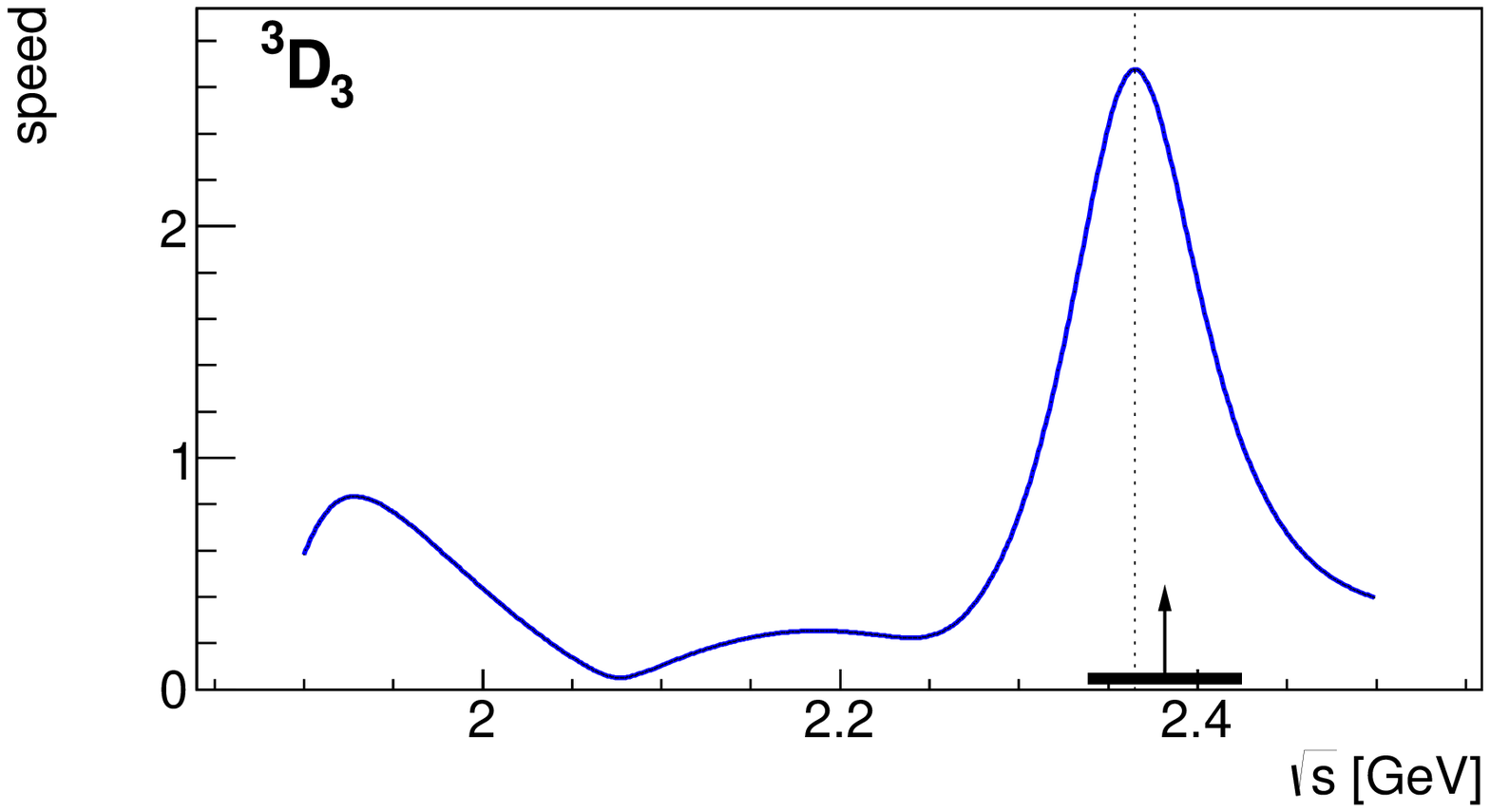} 
\caption{${\cal D}_{03}(2370)$ $\Delta\Delta$ dibaryon resonance signals 
in the Argand diagram on the left panel, and in the speed plot on the right 
panel, both for the $np$ $^3D_3$ partial wave, from recent $np$ scattering 
measurements by WASA-at-COSY \cite{wasa14}. 
Figures courtesy of Heinz Clement.} 
\label{fig:newdata} 
\end{figure*}

Further evidence supporting the ${\cal D}_{03}(2370)$ dibaryon assignment 
comes from very recent measurements of $pn$ elastic scattering as a function 
of energy, taking sufficiently small steps around $\sqrt{s}=2370$~MeV 
\cite{wasa14}. This is shown in Fig.~\ref{fig:newdata}--left for the Argand 
diagram of the $^3D_3$ partial wave, and in the right panel for the speed 
plot of the $^3D_3$ partial wave, within a new SAID partial wave analysis 
incorporating these measurements. 

$N\Delta$ and $\Delta\Delta$ $s$-wave dibaryon resonances ${\cal D}_{IS}$ 
with isospin $I$ and spin $S$ were proposed by Dyson and Xuong \cite{dyson64} 
as early as 1964, when quarks were still perceived as merely mathematical 
entities. They focused on the lowest-dimension SU(6) multiplet in the 
$\bf{56\times 56}$ product that contains the SU(3) $\overline{\bf 10}$ and 
${\bf 27}$ multiplets in which the deuteron ${\cal D}_{01}$ and $NN$ virtual 
state ${\cal D}_{10}$ are classified. This yields two dibaryon candidates, 
${\cal D}_{12}$ ($N\Delta$) and ${\cal D}_{03}$ ($\Delta\Delta$) as listed 
in Table~\ref{tab:dyson}. Identifying the constant $A$ in the resulting mass 
formula $M=A+B[I(I+1)+S(S+1)-2]$ with the $NN$ threshold mass 1878~MeV, a 
value $B\approx 47$~MeV was reached by assigning ${\cal D}_{12}$ to the $pp
\leftrightarrow \pi^+ d$ resonance at $\sqrt{s}=2160$~MeV (near the $N\Delta$ 
threshold) which was observed already during the 1950's. This led to the 
prediction $M({\cal D}_{03})$=2350~MeV. The ${\cal D}_{03}$ dibaryon has 
been the subject of several quark-based model calculations since 1980, 
see Ref.~\cite{D03calcs} for a representative although perhaps somewhat 
incomplete listing. 

\begin{table}[hbt]
\caption{Nonstrange s-wave dibaryon SU(6) predictions \cite{dyson64}.} 
\begin{tabular}{cccccc}
\hline\noalign{\smallskip} 
dibaryon & $I$ & $S$ & SU(3) & legend & mass \\
\noalign{\smallskip}\hline\noalign{\smallskip} 
${\cal D}_{01}$ & 0 & 1 & $\overline {\bf 10}$ & deuteron & $A$ \\ 
${\cal D}_{10}$ & 1 & 0 & ${\bf 27}$ & $nn$ & $A$ \\ 
${\cal D}_{12}$ & 1 & 2 & ${\bf 27}$ & $N\Delta$ & $A+6B$ \\ 
${\cal D}_{21}$ & 2 & 1 & ${\bf 35}$ & $N\Delta$ & $A+6B$ \\ 
${\cal D}_{03}$ & 0 & 3 & $\overline {\bf 10}$ & $\Delta\Delta$ & $A+10B$ \\ 
${\cal D}_{30}$ & 3 & 0 & ${\bf 28}$ & $\Delta\Delta$ & $A+10B$ \\ 
\noalign{\smallskip}\hline 
\end{tabular} 
\label{tab:dyson} 
\end{table} 

It is shown below that the pion-assisted methodology applied recently by Gal 
and Garcilazo \cite{gg13,gg14} couples ${\cal D}_{12}$ and ${\cal D}_{03}$ 
dynamically in a perfectly natural way, the analogue of which has not 
emerged in quark-based models. These hadronic-based calculations emphasize 
the long-range physics aspects of nonstrange dibaryons. Extensions to 
strangeness $\cal S$=$-1$ pion-assisted dibaryons are also briefly discussed.

\section{Pion-assisted nonstrange dibaryons} 


\subsection{$N\Delta$ dibaryons} 



\begin{figure*}[hbt]  
\includegraphics[width=0.8\textwidth]{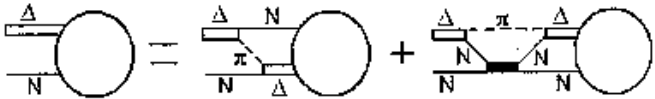} 
\caption{$N\Delta$ dibaryon's Lippmann-Schwinger equation \cite{gg14}.} 
\label{fig:DIS} 
\end{figure*}

The ${\cal D}_{IS}$ dibaryon candidates from Table~\ref{tab:dyson} have been 
calculated recently in Ref.~\cite{gg14} by solving Faddeev equations with 
relativistic kinematics for the $\pi NN$ three-body system, where the $\pi N$ 
subsystem is dominated by the $P_{33}$ $\Delta$(1232) resonance channel 
and the $NN$ subsystem is dominated by the $^3S_1$ and $^1S_0$ channels. 
The coupled Faddeev equations give rise then to an effective $N\Delta$ 
Lippmann-Schwinger (LS) equation for the three-body $S$-matrix pole, with 
energy-dependent kernels that incorporate spectator-hadron propagators, 
as shown diagrammatically in Fig.~\ref{fig:DIS} where circles denote the 
$N\Delta$ $T$ matrix. 

\begin{figure*}[thb]  
\includegraphics[width=0.48\textwidth,height=6.5cm]{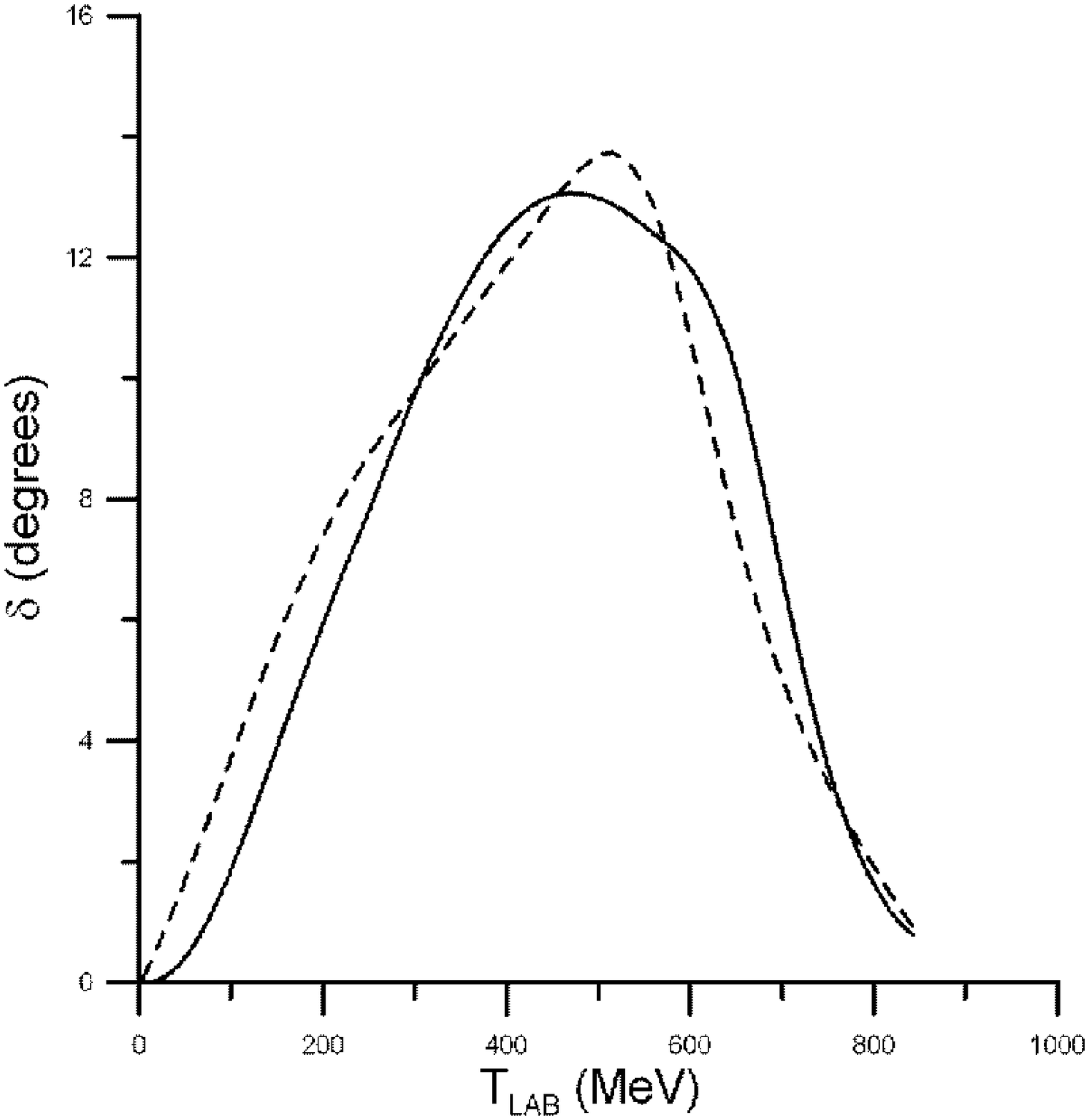} 
\includegraphics[width=0.48\textwidth,height=6.5cm]{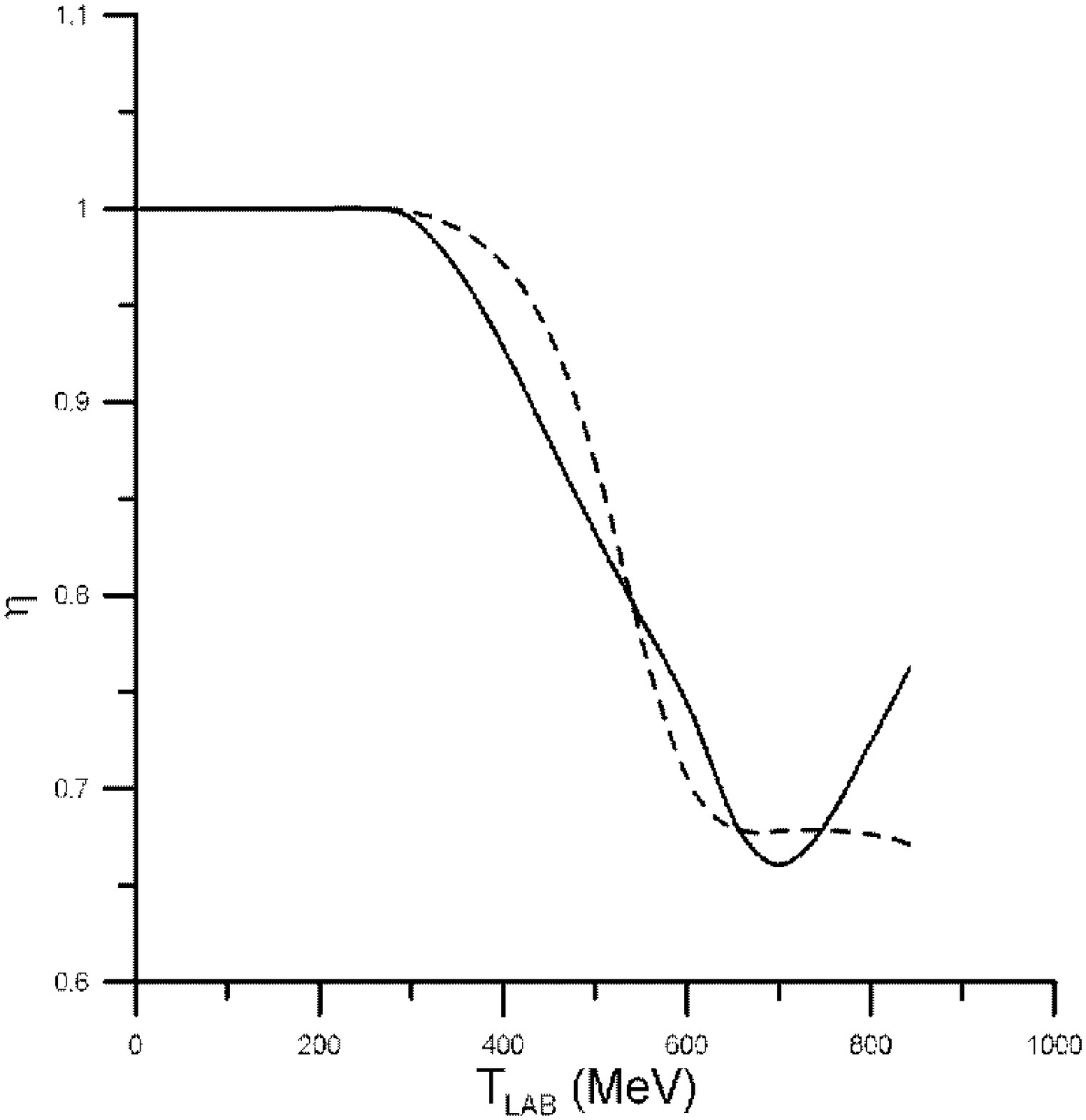} 
\caption{Coupled-channel fits (solid) to the SAID (dashed) $NN$ $^1D_2$ phase 
shift $\delta$ (left panel) and inelasticity $\eta$ (right panel) as obtained 
in Ref.~\cite{gg13}, see text.} 
\label{fig:fit} 
\end{figure*} 

Of the $L$=0 $N\Delta$ dibaryon candidates ${\cal D}_{IS}$ with 
$IS$=12,21,11,22, the latter two do not provide resonant solutions. 
For ${\cal D}_{12}$, only $^3S_1$ contributes out of the two $NN$ 
interactions, while for ${\cal D}_{21}$ only $^1S_0$ contributes. 
Since the $^3S_1$ interaction is the more attractive one, ${\cal D}_{12}$ 
lies below ${\cal D}_{21}$ as borne out by the calculated masses listed 
in Table~\ref{tab:NDel} for two choices of the $P_{33}$ interaction form 
factor corresponding to spatial sizes of 1.35~fm and 0.9~fm of the $\Delta$ 
isobar. The two dibaryons are found to be degenerate to within less than 
20~MeV, close to the $N\Delta$ threshold at $\approx$2.17~GeV with a width 
similar to that of the $\Delta$ baryon. In particular, the mass values 
calculated for ${\cal D}_{12}$ are reasonably close to the values 
$W=2148-{\rm i}63$~MeV \cite{arndt87} and $W=2144-{\rm i}55$~MeV \cite{hosh92} 
derived in $pp({^1D_2})$$\leftrightarrow$$\pi d({^3P_2})$ coupled-channel 
phenomenological analyses. 

\begin{table}[hbt] 
\caption{$N\Delta$ dibaryon $S$-matrix poles (in MeV) for ${\cal D}_{12}$ and 
${\cal D}_{21}$ obtained in Ref.~\cite{gg14} by solving $\pi NN$ Faddeev 
equations for two choices of the $\pi N$ $P_{33}$ form factor, with large 
(small) spatial size denoted $>$ ($<$).} 
\begin{tabular}{ccccc} 
\hline\noalign{\smallskip} 
$W^{>}({\cal D}_{12})$ & $W^{>}({\cal D}_{21})$ &  &
$W^{<}({\cal D}_{12})$ & $W^{<}({\cal D}_{21})$   \\  
\noalign{\smallskip}\hline\noalign{\smallskip} 
2147$-{\rm i}$60 & 2165$-{\rm i}$64 &  &  2159$-{\rm i}$70 & 
2169$-{\rm i}$69 \\
\noalign{\smallskip}\hline 
\end{tabular} 
\label{tab:NDel} 
\end{table} 

\subsection{$\Delta\Delta$ dibaryons} 

Generally, four-body $\pi\pi NN$ configurations appear in $\Delta\Delta$ 
dibaryons. Nevertheless, attempting to capture its most relevant degrees 
of freedom, the ${\cal D}_{03}$ dibaryon was studied in Ref.~\cite{gg13} 
by solving a $\pi N\Delta'$ three-body model, where $\Delta'$ is a stable 
$\Delta$(1232) and the $N\Delta'$ interaction is dominated by the 
${\cal D}_{12}$ dibaryon. The $I(J^P)=1(2^+)$ $N\Delta'$ interaction 
was not assumed to resonate but, rather, it was fitted within a 
$NN$--$\pi NN$--$N\Delta'$ coupled-channel caricature model to the 
$NN$ $^1D_2$ $T$-matrix, requiring that the resulting $N\Delta'$ 
separable-interaction form factor is representative of long-range physics, 
with momentum-space soft cutoff $\Lambda\leq 3$~fm$^{-1}$. A fit of this 
kind is shown in Fig.~\ref{fig:fit}. 

\begin{figure*}[htb]  
\includegraphics[width=0.7\textwidth]{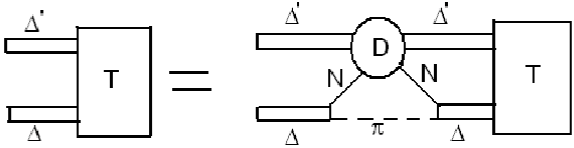} 
\caption{$S$-matrix pole equation for ${\cal D}_{03}(2370)$ $\Delta\Delta$ 
dibaryon \cite{gg13}.} 
\label{fig:D03} 
\end{figure*} 

The Faddeev equations of the $\pi N\Delta'$ three-body model give rise, 
as before, to an effective LS equation for the $\Delta\Delta'$ $S$-matrix pole 
corresponding to ${\cal D}_{03}$. This LS equation is shown diagrammatically 
in Fig.~\ref{fig:D03}, where $D$ stands for the ${\cal D}_{12}$ dibaryon. 
The $\pi N$ interaction was assumed again to be dominated by the $P_{33}$ 
$\Delta$ resonance, using two different parametrizations of its form 
factor that span a reasonable range of the $\Delta$ hadronic size. 
The calculation of ${\cal D}_{03}$ was extended in Ref.~\cite{gg14} 
to other ${\cal D}_{IS}$ $\Delta\Delta$ dibaryon candidates, with $D$ now 
standing for both $N\Delta$ dibaryons ${\cal D}_{12}$ and ${\cal D}_{21}$. 
Since ${\cal D}_{21}$ is almost degenerate with ${\cal D}_{12}$, and 
with no $NN$ observables to constrain the input $(I,S)$=(2,1) $N\Delta'$ 
interaction, the latter was taken the same as for $(I,S)$=(1,2). 
The lowest and also narrowest $\Delta\Delta$ dibaryons found are 
${\cal D}_{03}$ and ${\cal D}_{30}$. 

Representative results for ${\cal D}_{03}$ and ${\cal D}_{30}$ are assembled 
in Table~\ref{tab:DelDel}, where the calculated mass and width values listed 
in each row correspond to the specific spectator-$\Delta'$ complex mass 
$W(\Delta')$=1211$-{\rm i}x$49.5~MeV value used in the propagator of the LS 
equation shown in Fig.~\ref{fig:D03}. The value $x=1$ in the first row 
corresponds to the free-space $\Delta$(1232) $S$-matrix pole. It is implicitly 
assumed thereby that the decay $\Delta'\to N\pi$ proceeds independently of the 
$\Delta \to N\pi$ isobar decay. However, as pointed out in Ref.~\cite{gg13}, 
care must be exercised to ensure that the decay nucleons and pions satisfy 
Fermi-Dirac and Bose-Einstein statistics requirements, respectively. Assuming 
$L=0$ for the decay-nucleon pair, this leads to the suppression factor $x$=2/3 
depicted in the second row. It is seen that the widths obtained upon applying 
this width-suppression are only moderately smaller, by less than 15 MeV, than 
those calculated disregarding this quantum-statistics correlation. 

\begin{table}[hbt] 
\caption{$\Delta\Delta$ dibaryon $S$-matrix poles (in MeV) obtained in 
Refs.~\cite{gg13,gg14} by using in the propagator of the LS equation 
depicted in Fig.~\ref{fig:D03} a spectator-$\Delta'$ complex mass 
$W(\Delta')$=1211$-{\rm i}x$49.5~MeV, where $x$ is a width-suppression 
factor (see text). The last two columns give mass and width values averaged 
over those from the $>$ and $<$ columns, with $>$ and $<$ defined in 
Table~\ref{tab:NDel} caption. Other $\Delta\Delta$ dibaryon candidates are 
discussed in Ref.~\cite{gg14}.} 
\begin{tabular}{ccccccc} 
\hline\noalign{\smallskip} 
$x$ & $W^{>}({\cal D}_{03})$ & 
$W^{>}({\cal D}_{30})$ & 
$W^{<}({\cal D}_{03})$ & $W^{<}({\cal D}_{30})$ & $W_{\rm av}({\cal D}_{03})$ 
& $W_{\rm av}({\cal D}_{30})$  \\  
\noalign{\smallskip}\hline\noalign{\smallskip} 
1 & 2383$-{\rm i}$47 & 2412$-{\rm i}$49 & 
2342$-{\rm i}$31 & 2370$-{\rm i}$30 & 2363$-{\rm i}$39 & 2391$-{\rm i}$39 \\
$\frac{2}{3}$ & 2383$-{\rm i}$41 & 2411$-{\rm i}$41 & 
2343$-{\rm i}$24 & 2370$-{\rm i}$22 & 2363$-{\rm i}$33 & 2390$-{\rm i}$32 \\
\noalign{\smallskip}\hline 
\end{tabular} 
\label{tab:DelDel} 
\end{table} 

The mass and width values calculated for ${\cal D}_{03}$ \cite{gg13} 
agree very well with those determined by the WASA-at-COSY Collaboration 
\cite{wasa11,wasa13,wasa14}, reproducing in particular the reported width 
value $\Gamma({\cal D}_{03})\approx 70$~MeV which is considerably below the 
phase-space estimate $\Gamma_{\Delta}\leq\Gamma({\cal D}_{03})\leq 2\Gamma_{
\Delta}$, with $\Gamma_{\Delta}\approx 118$~MeV. No other calculation 
so far has succeeded to do that. Similarly small widths hold for 
${\cal D}_{30}$ which is located according to Table~\ref{tab:DelDel} about 
30 MeV above ${\cal D}_{03}$. Adding $\approx$20~MeV for the ${\cal D}_{21}$ 
input mass excess relative to ${\cal D}_{12}$, the resulting ${\cal D}_{30}$ 
to ${\cal D}_{03}$ mass excess of roughly 50~MeV agrees with that found 
recently by H.~Huang {\it et al.} \cite{D03calcs} in a quark-based 
calculation. A more complete discussion of these and of other ${\cal D}_{IS}$ 
$\Delta\Delta$ dibaryon candidates is found in Ref.~\cite{gg14}. 

Bashkanov, Brodsky and Clement \cite{BBC13} have emphasized recently the 
dominant role that six-quark hidden-color configurations might play in 
binding ${\cal D}_{03}$ and the exotic $I=3$ ${\cal D}_{30}$. The recent 
calculations by H.~Huang {\it et al.} \cite{D03calcs}, however, find that 
these configurations play a marginal role, enhancing dibaryon binding by 
merely 15$\pm$5~MeV and reducing the dibaryon width from 175 to 150 MeV 
for ${\cal D}_{03}$, still twice as big as the reported width, and from 
216 to 200 MeV for ${\cal D}_{30}$. These minor contributions of six-quark 
hidden-color configurations are in line with the secondary role found 
for them in studies of the $NN$ interaction in the context of the 
${\cal D}_{01}$ and ${\cal D}_{10}$ $NN$ `dibaryons' \cite{oka82}.

\section{Extension to strangeness ${\cal S}=-1$} 

\begin{figure*}[htb]
\includegraphics[width=0.7\textwidth]{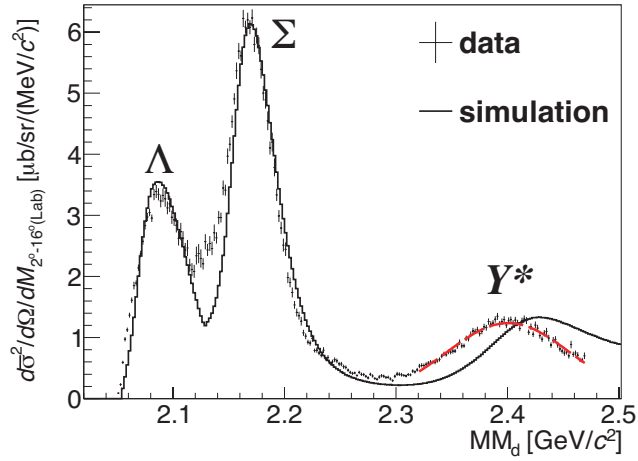}
\caption{J-PARC E27 missing-mass spectrum in $d(\pi^+,K^+)$ 
at 1.69 GeV/c \cite{E27a}. Figure courtesy of Tomofumi Nagae.} 
\label{fig:Y*N}
\end{figure*}

Recent searches of a $\Lambda(1405)N$ dibaryon have been reported 
from experiments at Frascati \cite{finuda13}, SPring-8 \cite{leps14}, 
GSI \cite{gsi14a} and J-PARC \cite{E15,E27a,E27b}. A missing-mass 
spectrum measured in the $d(\pi^+,K^+)$ reaction at 1.69 GeV/c in 
J-PARC is shown in Fig.~\ref{fig:Y*N}, indicating $\approx$22~MeV 
attractive shift of the unresolved $Y^{\ast}(1385+1405)$ quasi-free 
peak complex. This is consistent with the attraction expected in 
the $I=1/2,J^P=0^-$ $\Lambda(1405)N$ $s$-wave channel shown in 
Ref.~\cite{oka11} to overlap substantially with a $\bar KNN$ quasibound 
state known also as `$K^-pp$' which is being searched for in these 
experiments. The lower-energy components of this $\bar KNN$ 
dibaryon--$\pi\Lambda N$ and $\pi\Sigma N$--do not support 
any strongly attractive meson-baryon $s$-wave interaction. 

The $\pi\Lambda N$--$\pi\Sigma N$ system, however, can benefit from strong 
meson-baryon $p$-wave interactions fitted to the $\Delta(1232)\to\pi N$ 
and $\Sigma(1385)\to\pi\Lambda$--$\pi\Sigma$ form factors by fully aligning 
isospin and angular momentum: $I=3/2$, $J^P=2^+$. 
Such ${\cal S}=-1$ pion-assisted dibaryon was introduced in Ref.~\cite{gg08}, 
predicting a dibaryon resonance about 10--20 MeV below the $\pi\Sigma N$ 
threshold obtained by solving $\pi YN$ coupled-channel Faddeev equations 
\cite{gg13a}. This prediction, however, is sensitive to the $p$-wave form 
factors assumed. Adding a $\bar KNN$ channel hardly matters, since its 
leading $^3S_1$ $NN$ configuration is Pauli forbidden. 

This ${\cal S}=-1$ pion-assisted dibaryon, denoted $\cal Y$, overlaps with 
$s$-wave $^5S_2$, $I=3/2$ $\Sigma(1385)N$ and $\Delta(1232)Y$ dibaryon 
configurations, the lower of which is $\Sigma(1385)N$. These quantum numbers 
differ from $^1S_0$, $I=1/2$ for $\Lambda(1405)N$ which is being searched 
upon. A recent search for the $I=3/2$ $\cal Y$ dibaryon in 
\begin{eqnarray} 
   p~ + ~p & ~\rightarrow ~ & {\cal Y}^{++} ~+~K^0 \nonumber  \\  
           &                & ~\hookrightarrow ~ \Sigma^+ ~+~ p \; 
\label{eq:pptoY++} 
\end{eqnarray} 
by the HADES Collaboration at GSI \cite{gsi14b} found no $\cal Y$ dibaryon 
signal. 
Other possible search reactions are  
\begin{eqnarray} 
\pi^{\pm} ~+~ d & ~\rightarrow ~ & {\cal Y}^{++/-} ~+~K^{0/+}  \nonumber  \\ 
 &  &  ~\hookrightarrow ~ \Sigma^{\pm} +p(n) \; , 
\label{eq:pi+dtoY++} 
\end{eqnarray} 
again offering distinct $I=3/2$ decay channels. Other decay channels such as 
\begin{eqnarray} 
\pi^+ ~+~ d & ~\rightarrow ~ & {\cal Y}^+ ~+~K^+  \nonumber  \\ 
 &  &  ~\hookrightarrow ~ \Sigma^0 + p  
\label{eq:pi+dtoY+} 
\end{eqnarray} 
allow for both $I=1/2,3/2$. E27 has just reported \cite{E27b} a dibaryon 
signal near the $\pi\Sigma N$ threshold in reaction (\ref{eq:pi+dtoY+}). 
This requires further experimental study.

\section{Conclusion} 

It was shown how the 1964 Dyson-Xuong SU(6)-based classification and 
predictions of nonstrange dibaryons \cite{dyson64} are confirmed in the 
hadronic model of $N\Delta$ and $\Delta\Delta$ pion-assisted dibaryons 
\cite{gg13,gg14}. The input for dibaryon calculations in this model consists 
of nucleons, pions and $\Delta$'s, interacting via long-range pairwise 
interactions. These calculations reproduce the two nonstrange dibaryons 
established experimentally and phenomenologically so far, the $N\Delta$ 
dibaryon ${\cal D}_{12}$ \cite{arndt87,hosh92} and the $\Delta\Delta$ 
dibaryon ${\cal D}_{03}$ \cite{wasa11,wasa13,wasa14}, and predict several 
exotic $N\Delta$ and $\Delta\Delta$ dibaryons. We note that, within the 
$\pi N\Delta$ three-body model of ${\cal D}_{03}$, ${\cal D}_{12}$ provides 
a two-body decay channel $\pi {\cal D}_{12}$ with threshold lower than 
$\Delta\Delta$ which proves instrumental in obtaining a relatively small 
width for ${\cal D}_{03}$ \cite{gg14}.  

Finally, a straightforward extension of nonstrange pion-assisted dibaryon 
phenomenology to strangeness $\cal S$=$-1$ was briefly discussed in connection 
to recent searches of kaonic nuclear clusters, see Ref.~\cite{gal13} for 
a recent review.

\begin{acknowledgements} 
Fruitful collaboration with Humberto Garcilazo and stimulating discussions 
with Mikhail Bashkanov, Heinz Clement, Tomofumi Nagae and Makoto Oka are 
gratefully acknowledged. Special thanks are due to the Organizers of EXA14 
for their kind hospitality. Support by the EU FP7 initiative HadronPhysics3, 
under the SPHERE and LEANNIS cooperation programs, is gratefully acknowledged. 
\end{acknowledgements}


\begin{thebibliography}{99}

\bibitem{wasa11} P.~Adlarson et al. (WASA-at-COSY Collaboration), Phys. Rev. 
Lett. {\bf 106} (2011) 242302, and arXiv:1409.2659. See also the preceding 
reports: H.~Clement et al. (CELSIUS-WASA Collaboration), Prog. Part. Nucl. 
Phys. {\bf 61} (2008) 276; M.~Bashkanov et al. (CELSIUS/WASA Collaboration), 
Phys. Rev. Lett. {\bf 102} (2009) 052301. 

\bibitem{wasa13} P.~Adlarson et al. (WASA-at-COSY Collaboration), 
Phys. Lett. B {\bf 721} (2013) 229. 

\bibitem{wasa14} P.~Adlarson et al. (WASA-at-COSY Collaboration, SAID 
Data Analysis Center), Phys. Rev. C {\bf 90} (2014) 035204. See also 
P.~Adlarson et al. (WASA-at-COSY Collaboration, SAID Data Analysis Center), 
Phys. Rev. Lett. {\bf 112} (2014) 202301. 




\bibitem{dyson64} F.J.~Dyson, N.-H.~Xuong, Phys. Rev. Lett. {\bf 13} (1964) 
815. 

\bibitem{D03calcs} P.J.~Mulders, A.T.~Aerts, J.J.~de Swart, Phys. Rev. D 
{\bf 21} (1980) 2653; M.~Oka, K.~Yazaki, Phys. Lett. B {\bf 90} (1980) 41; 
M.~Cveti\v{c}, B.~Golli, N.~Manko\v{c}-Bor\v{s}tnik, M.~Rosina, Phys. Lett. 
B {\bf 93} (1980) 489; P.J.~Mulders, A.W.~Thomas, J. Phys. G {\bf 9} (1983) 
1159; K.~Maltman, Nucl. Phys. A {\bf 438} (1985) 669; T.~Goldman, K.~Maltman, 
G.J.~Stephenson, K.E.~Schmidt, F.~Wang, Phys. Rev. C {\bf 39} (1989) 1889; 
X.Q.~Yuan, Z.Y.~Zhang, Y.W.~Yu, P.N.~Shen, Phys. Rev. C {\bf 60} (1999) 
045203; R.D.~Mota, A.~Valcarce, F.~Fern\'{a}ndez, D.R.~Entem, H.~Garcilazo, 
Phys. Rev. C {\bf 65} (2002) 034006; J.L.~Ping, H.X.~Huang, H.R.~Pang, 
F.~Wang, C.W.~Wong, Phys. Rev. C {\bf 79} (2009) 024001. H.~Huang, J.~Ping, 
F.~Wang, Phys. Rev. C {\bf 89} (2014) 034001. 











\bibitem{gg13} A.~Gal, H.~Garcilazo, Phys. Rev. Lett. {\bf 111} (2013) 172301. 

\bibitem{gg14} A.~Gal, H.~Garcilazo, Nucl. Phys. A {\bf 928} (2014) 73.  

\bibitem{arndt87} R.A.~Arndt, J.S.~Hyslop III, L.D.~Roper, Phys. Rev. D 
{\bf 35} (1987) 128. 

\bibitem{hosh92} N.~Hoshizaki, Phys. Rev. C {\bf 45} (1992) R1424, 
Prog. Theor. Phys. {\bf 89} (1993) 563. 




\bibitem{BBC13} M.~Bashkanov, S.J.~Brodsky, H.~Clement, Phys. Lett. 
B {\bf 727} (2013) 438. See also F.~Huang, Z.Y.~Zhang, P.N.~Shen, W.L.~Wang, 
arXiv:1408.0458 [nucl-th]. 

\bibitem{oka82} S.~Ohta, M.~Oka, A.~Arima, K.~Yazaki, Phys. Lett. B {\bf 119} 
(1982) 35. 


\bibitem{finuda13} M.~Agnello et al. (FINUDA Collaboration), Nucl. Phys. A 
{\bf 914} (2013) 310. 

\bibitem{leps14} A.O.~Tokiyasu et al. (LEPS Collaboration), Phys. Lett. B 
{\bf 728} (2014) 616. 

\bibitem{gsi14a} G.~Agakishiev et al. (HADES Collaboration), Phys. Lett. B 
{\bf 742} (2015) 242. 

\bibitem{E15} T.~Hashimoto et al. (J-PARC E15 Experiment), arXiv:1408.5637. 

\bibitem{E27a} Y.~Ichikawa et al. (J-PARC E27 Experiment), Prog. Theor. Exp. 
Phys. {\bf 2014}, 101D03. 

\bibitem{E27b} Y.~Ichikawa et al. (J-PARC E27 Experiment), Prog. Theor. Exp. 
Phys. {\bf 2015}, 021D01. 

\bibitem{oka11} T.~Uchino, T.~Hyodo, M.~Oka, Nucl. Phys. A {\bf 868-869} 
(2011) 53. 

\bibitem{gg08} A.~Gal, H.~Garcilazo, Phys. Rev. D {\bf 78} (2008) 014013, 
Phys. Rev. C {\bf 81} (2010) 055205. 

\bibitem{gg13a} H.~Garcilazo, A.~Gal, Nucl. Phys. A {\bf 897} (2013) 167. 

\bibitem{gsi14b} J.C.~Berger-Chen, L.~Fabbietti, in Proc. PANIC 2014, 
arXiv:1410.8004.  

\bibitem{gal13} A.~Gal, Nucl. Phys. A {\bf 914} (2013) 270. 

\end{thebibliography}
\end{document}